\documentclass[sn-mathphys]{sn-jnl}
\usepackage{graphicx, listings, setspace, verbatim}
\usepackage{amsmath, amssymb, mathtools, esint}

\usepackage[utf8]{inputenc}

\usepackage{amsmath, amssymb, mathtools, esint}

\jyear{2021}%

\begin{document}

\title{An Epistemic Analysis of Time Phenomenon}


\author*[1]{\fnm{Farhang} \sur{Hadad Farshi}}\email{farhang.hadad@runbox.com}

\author*[1,2]{\fnm{Silvia} \sur{DeBianchi}}\email{
	Silvia.DeBianchi@unimi.it}

\affil*[1]{\orgdiv{Department of Philosophy}, \orgname{Universidad Aut\'{o}noma de Barcelona}, \city{Barcelona} \postcode{08193}, \country{Spain}}

\affil[2]{\orgdiv{Department of Philosophy}, \orgname{Universit\`{a}	 degli Studi di Milano},  \city{Milan} \postcode{20122}, \country{Italy}}


\abstract{In this work we present an epistemic analysis of time phenomenon using the mathematical machinery of information theory and modular theory. By adopting limited commitment to the ontology of time evolution, and instead by mainly relying on the information that is in principle accessible to the observer, we find that the most primary aspect of the temporal experience, the perceived distinctiveness across the states of the world, emerges as a purely epistemic function. By analyzing the mathematical properties of this epistemic function, we interpret it to be in principle insensitive to any ontic state of the world, which leads to the conclusion that the observer is subject to temporal experience irrespective of whether the underlying state of the world is dynamical or invariant. On the ground of the presented analysis, we also provide a solution to the conceptual challenge of non-equilibrium phenomena that faces the thermal time hypothesis.}




\maketitle

\section{Introduction}

	What does the phenomenon of time consist of? Within the phenomenology of time \citep{Hussy, Soko, Kelly}, the most primary aspect of the temporal experience  is viewed to be constituted by the perceived change in the state of affairs, the distinctiveness of the states of a system from one moment to another, in the eyes of the observer. An obvious, yet central point is that for the notion of change to make any sense at all, the observer should be able to distinguish the states of the observed system from one moment to the next by contrasting them against each other. Therefore, the apprehension of time becomes possible in the observer's capacity to distinguish the states of affairs, that are distinct but not isolated, through observing the world around themselves {\textit{i.e.}} through performing measurements. Our contribution focuses on the way in which we understand the change or the dynamics from the perspective of an observer. In physics however, one understands the change or the dynamics of a system in terms of a time-evolution operator that drives the system from one state to another. For instance, in quantum theory, the state of a system changes according to some unitary transform which might have been generated internally or externally to that system. Furthermore, the restriction of this unitary transform to subsystems can induce a change with some direction {\textit{e.g.}} tendency towards disorder.\\

\noindent The correspondence between  these two notions of change, one due to the physical evolution encoded in dynamical operators, and the other due to the observer's perception of variation  appears quite blurry. More concretely, is it possible for the world to be in an invariant state, but nevertheless to appear as dynamical? Or can the behaviour of a subsystem that is governed by an entropy-decreasing transform appear to be tending towards disorder? In other words, how big is this apparent gap between an ontic notion and an epistemic notion of change?\\

\noindent This work is an attempt to explicate the phenomenon of time, which adopts limited commitments to the ontology of time evolution, and instead takes on an epistemological attitude, and relies solely on the information that is in principle accessible to the observer. Here we subscribe to the view  \citep{Witty,Williams} that any epistemic inquiry takes place within a context that itself is immune from skeptical assessments. That is, in each context there exists a set of commitments or `hinges' on the ground of which the epistemic practice becomes possible. In this work the hinge consists of the belief that our world admits  a local and unitary structure that is depicted \citep{haag} by the language of type III von Neumann algebra \citep{dix, connesNG} , of which an observer can in principle obtain or construct information through local measurements. \\

\noindent Our goal thus is to provide an epistemic analysis of the time phenomenon from within the framework of type III von Neumann algebra and information theory. Type III structure implies two important properties that will play a crucial role in our analysis. The first property is the unavoidable presence of entanglement across regions of space time and their causal complements \citep{RevModPhys.90.045003}. The second property is that the type III structures are intrinsically dynamical, a result that as has been analysed and understood using the mathematical machinery of modular theory \citep{TT}. We will see that in this approach, time phenomenon emerges as an epistemic function with the property that it is in principle insensitive to any particular ontic state of the world, and consequently insensitive with respect to any ontic evolution of the world as all the other global states induce the same phenomenon. On the ground of this analysis, we will also be able to provide a solution to one of the conceptual problems that is linked with the Thermal Time Hypothesis \citep{connes}, which has been recently voiced \citep{swanson}. Thermal time hypothesis as a solution to the problem of time in generally covariant quantum theories, is the proposition that time has a thermodynamical origin. That is, based on the modular theory of von Neumann algebra, any state defines a mathematical transform with respect to which it is in equilibrium, and this transform {\textit{is}} the experienced time flow. The problem with this claim is that it does not seem to address the highly non-equilibrium phenomena that we perceive around us since the state of the world that determines the time flow remains at thermal equilibrium with respect to this flow. We will see how does this conceptual issue can be dissolved within the analysis developed in section 4.\\

\noindent In section 2, we review the concept of relative entropy and distinguishability in information theory, and discuss the key result of monotonicity. In section 3, we provide a brief introduction to the type III von Neumann algebra, and discuss the two central concepts of entanglement and modularity and their operational interpretations, which apply to this structure.  In section 4, we present an information-theoretic definition of change, and by analysing its algebraic properties we discuss some of the key conceptual implications in regards to the emergence of the time phenomenon. In section 5, we address the conceptual issue of non-equilibrium phenomena, which is linked with the thermal time hypothesis, and argue how this problem is dissolved within the approach that we pursue. Finally, in section 6, we conclude by reflecting on the potential directions in which the presented analysis can be furthered.

\section{Relative Entropy: A Key Concept}

	Since the connection between the perception of change in the state of a system and the capacity to distinguish the system's distinct states plays an essential role in our analysis, we would like to have a  mathematically sharp understanding of distinguishability.\\

\noindent Central to a rigorous description of distinguishing is the measure of \textit{information} or \textit{surprise} \citep{shannon} that is associated with a probability distribution. Suppose that a certain event $x$ has an occurrence probability of $p_x$. If the event is highly probable, it will be little surprise when it happens, and thus the transmission of the message indicating that $x$ has occurred carries very little information. On the other hand, if $p_x$ is very small the occurrence of $x$ as a rare event is highly informative. Therefore, the information (surprise) content of an event $x$ is a function that increases as $p_x$ decreases, and can be expressed as $-\log p_x$ \footnote{The reason behind a logarithmic definition of information content of an even is the additivity of the amount of surprise associated with the occurrence of two independent events $x$ and $y$: $-\log p_{xy} = -\log p_x -\log p_y$.}. It can be easily seen that the average amount of surprise in a set of events $\{x\}$ with the corresponding probability distribution $\{p_x\}$, or alternatively, the information content of a message comprising the occurrence of a sequence of events $\{x\}$ with probabilities $\{p_x\}$ is

\begin{equation}
H(p) = -\sum_x p_x\log p_x. \nonumber
\end{equation}

\noindent $H(p)$ is called the (Shannon) entropy associated with $p_x$. Now suppose that the probabilities for a collection of events $\{x\}$ is $\{p_x\}$, but we erroneously believe it to be $\{q_x\}$. Based on this false belief, the surprise associated with a certain event $x$ is $-\log q_x$, and subsequently, the average surprise or the information content of a message comprising the occurrence of a sequence of events $\{x\}$ is $-\sum_xp_x\log q_x$ (since irrespective of our false belief the event $x$ occurs with probability $p_x$.) For instance, we typically assume that a coin to be fair $\textit{e.g.}$ if thrown many times, the distribution of heads and tails are equal, whereas in reality this assumption is never true, and after many trials of throwing the coin we become aware of a discrepancy between the expected and the observed outcomes. This discrepancy  that is given as the difference  between the average surprise associated with the expected probability distribution $q$ and the real probability distribution $p$:

\begin{equation}
H(p,q) = \sum_x p_x(\log p_x - \log q_x) \nonumber
\end{equation}

\noindent is called the relative entropy, a mathematical concept that was originally developed in cryptanalysis  \cite{kullback}, and it is a measure of how distinguishable are two probability distributions from one another, or alternatively, it is a measure of how likely it is to confuse the information contents of two distinct messages. For a pair of probability distributions $\{p , q\}$ that are indistinguishable the relative entropy $H(p,q) = 0$, and this is true if and only if $p=q$. In contrast, for a perfectly distinguishable pair, $H(p,q) = \infty$. To illustrate this, imagine that we are in possession of a fair coin but we are under the impression that the coin is completely unfair \textit{e.g.} the coin always comes up heads. In the course of throwing this coin few times, the coin will eventually land on tails, which leads to our immediate realisation that the held assumption was incorrect . That is, we are able to perfectly distinguish the expected distribution from the observed one. By computing the relative entropy for this coin example one obtains $H(p,q) = \infty$. On the other hand, if our erroneous belief was that the coin is partially unfair \textit{e.g.} the coin comes up heads third of the time, the corresponding relative entropy would have been finite. This implies the expected and the observed distributions are not perfectly distinguishable, and the only way for us to be able to gain full confidence in their distinction is to throw the coin infinite number of times.\\

\noindent Let us consider the situation where $\{x\}$ and $\{y\}$ are two collections of events with the joint probability distribution $p_{xy}$, and that we mistakenly believe that the joint distribution to be $q_{xy}$. After many trials, our confidence that our belief is false is determined by the relative entropy $H(p_{xy},q_{xy})$. But suppose that we only observe $\{x\}$ and not $\{y$\},  in which case our confidence in differentiating the believed distribution from the observed one is computed in terms of the relative entropy between the marginal probability distributions $p_x = \sum_y p_{xy}$ and $q_{x} = \sum_yq_{xy}$. The \textit{monotonicity} \citep{RevModPhys.74.197} of relative entropy implies that the distinguishability between the probability distributions of a system never increases as we restrict our observation to part of the system by integrating out the rest:
\begin{eqnarray}
H(p_{x},q_{x} )\leq H(p_{xy},q_{xy}). \nonumber
\end{eqnarray}
	\noindent The intuition behind the monotonicity is that our ability to distinguish two probability distributions always decays under coarse graining \citep{coarse}. That is, as our access to the distributions becomes less precise, the distinction between them becomes less visible.\\

\noindent Quantum mechanics can be viewed as the \textit{non-commutative} \citep{connesNG, QP} generalisation of the classical \textit{commutative} probability theory. This view of quantum theory allows us to generalise the analysis of distinctions across probability distributions in classical systems to distinctions across states in quantum systems. More concretely, in transitioning from the classical theory to the quantum theory, the commutative space of events with a corresponding probability distribution is generalised to the non-commutative space of \textit{observable algebras} with a corresponding \textit{state} that is a positive, trace-class linear operator on a Hilbert space with unit trace. The state $\rho$ defines a normalised positive linear functional over the set of observables $A$ as the average or the expectation value of $A$:
\begin{eqnarray}
\rho(A) =  \textrm{Tr}\rho A.
\label{state}
\end{eqnarray}
\noindent Similar to the classical probability theory, the  information or the surprise associated with a state $\rho$ is defined as $-\log \rho$, and using Eq.(\ref{state}), the average surprise or the information content of $\rho$ reads:
\begin{equation}
H(\rho) = \rho(-\log\rho) = -\textrm{Tr}\rho\log\rho. \nonumber
\end{equation} 
$H(\rho)$ is called the (von Neumann) entropy of the state $\rho$. In order to obtain a better understanding of entropy in quantum systems let us consider the example where $\rho$ describes the state of a two-level system $\{\vert\pm\rangle\}$. If $\rho$ is a pure state  and we are told what that state is, there will be zero surprise associated with the result of our observation (measurement). For instance, let us assume that the system is prepared in the ground state $\rho_p = \vert -\rangle\langle -\vert$ or in some superposition of ground and excited state $\rho_p = 1/2\{(\vert +\rangle +\vert -\rangle)(\langle +\vert +\langle -\vert )\}$, and we have been informed of these state preparations. One can easily check that the entropy for both of these states is zero. This means that the result of our measurement will always be in full agreement with our expectation, and thus our observation will be devoid of any uncertainty. Therefore, a pure state can be interpreted as a state of which perfect knowledge can be had, and that leads to the absence of any surprise, or alternatively, to the absence of any information conveyed upon observing it. This is analogous to the classical case where the event $x$ occurs with unit probability, and therefore the degree to which we are surprised after observing its occurrence is zero.\\

\noindent In contrast, for a system that is prepared in a mixed state, $\rho_m = 1/2\{\vert + \rangle\langle +\vert +\vert -\rangle\langle -\vert \}$, the entropy is non-zero, which implies that there will be a degree of uncertainty built into this system, and that inevitably leads to some degree of surprise in the result of our observation. That is, in principle for mixed states a perfect knowledge of the outcome of a measurement can not be had beforehand. An insightful way of interpreting a mixed state is to consider the system of interest to be sharing a pure entangled state with an ancillary system to which we do not have any access. In this view, the presence of surprise in our observation is rooted in our ignorance in regards to the pure state of the entirety of this bipartite system.   \\

\noindent Analogous to the classical systems, relative entropy in quantum systems is defined as the difference between the average surprise associated with the expected and the real state of that system:
\begin{equation}
H(\rho,\sigma) = \textrm{Tr} \rho(\log \rho - \log \sigma).
\label{RE}
\end{equation}

\noindent $H(\rho,\sigma)$ is measure of how well an observer can  distinguish the state $\rho$ from $\sigma$  through measurements. Finally, the monotonicity of relative entropy in quantum systems \citep{RevModPhys.74.197} implies that distinguishability never increases under \textit{restriction}:
\begin{eqnarray}
H(\rho_{A},\sigma_{A})\leq H(\rho_{AB},\sigma_{AB})
\label{monotony}
\end{eqnarray}
where $\rho_{AB}$ denotes the state of bipartite system $AB$ , and $\rho_A$ is the reduced state  obtained by restricting our access to the subsystem $A$, which can be mathematically implemented through a partial trace over the subsystem $B$: $\rho_A = \textrm{Tr}_{B}\rho_{AB}$. To illustrate the significance of Eq.(\ref{monotony}) let us consider both the expected and the real bipartite states to be entangled and perfectly distinguishable: 
\begin{eqnarray}
\rho_{AB} &=& \vert \psi\rangle\langle\psi\vert  \hspace{5mm}\textrm{with}\hspace{5mm} \vert \psi\rangle = \frac{1}{\sqrt{2}}\{\vert ++\rangle+\vert--\rangle\}\nonumber \\
\sigma_{AB} &=& \vert \phi\rangle\langle\phi\vert \hspace{5mm}\textrm{with}\hspace{6mm} \vert \phi\rangle = \frac{1}{\sqrt{2}}\{\vert -+\rangle+\vert +-\rangle\} \nonumber
\end{eqnarray}

\noindent Form the perspective of a global observation the bipartite states are perfectly distinguishable $H(\rho_{AB},\sigma_{AB}) = \infty$. In the eyes of a local observer who is restricted to either one part of the system however, the two states are indistinguishable $	H(\rho_{A},\sigma_{A}) = 0$. In other words, restriction to any part of a composite system in an entangled state leads to the loss of information that is shared across the global state, and this inevitably results in the decay in observers ability to tell states apart, hence the reduction in the relative entropy.
\section{The World in the Language of Type III Algebra}

In this section we will address the structure of the world in which the observer is situated, and the commitment to this structure functions as the hinge in our epistemic analysis. This hinge is the belief that the world or the context in which the observer is situated admits a local and unitary structure, which is depicted by the language of type III von Neumann algebra.\\

\noindent The phenomena in quantum regime are described in terms of the algebra of observables and their expectations, which is the non-commutative generalization of the commutative algebra of events and their expectations in classical probability theory. It turns out that the non-commutative probability theory can be rigorously modeled by the von Neumann algebra \citep{connesNG, QP}.\\ 

\noindent A von Neumann algebra $\mathcal{N}$ can be  construed as the $*$-algebra of bounded operators $\mathcal{B}(\mathcal{H})$ on a separable Hilbert space that is closed under weak limits. The term $*$-algebra simply means that if the operator $A$ lives in the algebra, so does its adjoint $A^*$, and moreover, the observables of the theory are represented by those operators that are self-adjoint: $A = A^{\ast}$. The term closure under weak limits addresses the boundaries of the algebra, and can be interpreted in the following way. If the expectation value of a sequence of operators $\{A_1,..., A_n\}\in\mathcal{N}$,  converges to the expectation value of some operator $A$ as $n$ grows, then $A$ is also included in the algebra. In other words, if the observer can not distinguish the operator $A$ from $A_n$ at large $n$ through \textit{measurements} (that are encoded in the expectation values of the operators), then from the perspective of this observer $A = \lim_{n\rightarrow\infty}A_n$. Therefore, a von Neumann algebra can be thought of as a formal language in terms of which the quantum phenomena, that appear to the observer through measurements, can rigorously be depicted.\\

\noindent In a sequence of papers \cite{VonNeumann, VonNeumannII, VonNeumannIII, VonNeumannIV} Murray and von Neumann investigated certain class of algebras called \textit{factors}. The idea behind a factor is to distill  the algebra $\mathcal{N}$ into separate commuting sub-algebras  $\mathcal{N} = \mathcal{R}\cup\mathcal{R}^{\prime}$ that only share trivial elements with each other:
\begin{eqnarray}
\mathcal{R} \cap \mathcal{R}^{\prime} = \mathbb{C}\mathbb{I}. \nonumber
\end{eqnarray}
The theory of factors was mainly motivated by the question regarding the divisibility of the world as a quantum system into separate independent subsystems. More concretely, the theory concerns the conditions under which an algebra describing a quantum system can be broken down into factors each describing the corresponding subsystem, and this lead to the classification of factors into  types: I, II, and III.\\

\noindent In the case of type I, the algebra is factored based on the tensor-product factorisation of the underlying Hilbert space of the system, $\mathcal{H} = \mathcal{H}_1\otimes\mathcal{H}_2$ . The observables of each subsystem belong to the algebra of bounded operators that act only on the Hilbert subspace corresponding to that subsystem: $\mathcal{R}_1 = \mathcal{B}(\mathcal{H}_1)\otimes \mathbb{I}$ and  $\mathcal{R}_2 = \mathbb{I}\otimes\mathcal{B}(\mathcal{H}_2)$. In this way the algebra of the entire system is factored :
\begin{eqnarray}
\mathcal{N} = \mathcal{R}_1\cup\mathcal{R}_2 = \mathcal{B}(\mathcal{H}_1)\otimes\mathcal{B}(\mathcal{H}_2). \nonumber
\end{eqnarray}
Type I factors depict quantum phenomena in a world that is not local, and in which systems can be broken down into individual subsystems. In other words, in a world in which observers do not perceive locality, quantum phenomena can be faithfully formalised in terms of the type I von Neumann algebra. \\

\noindent As locality enters the world of the observer, the type I picture of the world no longer holds. This has been illustrated by a rigorous analysis \cite{PhysRevLett.73.613} of the gedankenexperiment proposed by Fermi \cite{fermi}. The experiment consists of a pair of two-level system placed at some finite distant $d$ from each other, and where one system is in the excited state and the other in the ground state$,\{\vert +\rangle_1 , \vert -\rangle_2\}$ . If at some point the exited system decays by emitting a photon, then the probability that the second system will become excited should not change at least for the duration of $d/c$ where $c$ denotes the speed of light. However, by analysing the experiment within the type I structure, where the Hilbert space of the entire system is factored into separate individual subspaces, $\mathcal{H} = \mathcal{H}_1\otimes \mathcal{H}_2\otimes\mathcal{H}_3$, corresponding to the systems $1$, $2$, and the photon, one computes this probability to be nonzero\footnote{The  interaction Hamiltonian is bounded.} instantly after the excited system has decayed. More concretely, the excitation probability for the second system is encoded in the observable $P = \mathbb{I}_1\otimes \vert +\rangle_2\langle+\vert \otimes\mathbb{I}_3\in\mathcal{R}_2$, for which the expectation $\rho(P(t)) \neq 0$ for $t>0$. That is, the application of the type I picture in alaysing the gedankenexperiment  leads to the violation of locality as perceived. \\

\noindent Type III factors, on the other hand, depict a quantum mechanical world, in which systems have well defined localisation properties. In this picture, the Hilbert space of the world is not factored into separate, individual subspaces with corresponding observable algebras, but instead is \textit{enframed} by a \textit{net} of observable algebras. More specifically, in this picture, the von Neumann  algebra is factored by associating to each bounded region of space time a set of observables that encode the physical properties of the global system from the perspective of an observer located in that region. In other words, in the type III structure, the idea of subsystems with their own observable algebras is dissolved, and is replaced by the idea of bounded regions of space time with the corresponding observable algebras localised in those regions, and it is through these local algebras that the entire system (with its unbroken Hilbert space) is observed.\\

\noindent Therefore, in type III setting, the algebra of observables are labeled  by the region in which they reside: $\mathcal{R}_{\mathcal{O}} \equiv \mathcal{N}(\mathcal{O})$, where $\mathcal{O}$ denotes an arbitrary space time region. The operational interpretation of this labeling is that two distinct states $\{\rho , \sigma\}$ are indistinguishable from the perspective of the observer located in $\mathcal{O}$ when :
\begin{eqnarray}
\rho(A) = \sigma(A) \hspace{5mm} \textrm{for all} \hspace{5mm} A\in\mathcal{R}_{\mathcal{O}}. \nonumber
\end{eqnarray}
In other words, the observer who is situated in the region $\mathcal{O}$ observes the world by performing measurements on the observables at his disposal, $\mathcal{R}_{\mathcal{O}}$, and as long as the results of these measurements coincide for the two distinct states, the observer can not tell them apart. Within this frame, the gedankenexperiment can be rigorously formalised without leading to any violation of the local and causal structure of the space time as following. \\

\noindent The observable algebra $\mathcal{R}_i$ in type I frame is replaced with $\mathcal{R}_{\mathcal{O}_i}$ where $\mathcal{O}_i = t\times \mathcal{V}_i \in \mathbb{R}^4$, is the space time region - comprising the spatial volume $\mathcal{V}_i$ at time $t$ - in which the system $i$ is localised . In order to observe the effect of the first system's decay on the second system at any time $t$ , the observer who is located in $\mathcal{O}_2$ must contrast the global tripartite state $\rho$ (representing the case where the system $1$ has not decayed) against the state  $\sigma$ (representing the case where the system $1$ has decayed a photon). These two states are indistinguishable in the eyes of this observer who has access only to  the observables $A\in\mathcal{R}_{\mathcal{O}_2}$ for the duration $t<d/c$ where $d$ denotes the spatial separation between the regions $\mathcal{V}_1$ and $\mathcal{V}_2$ :
\begin{eqnarray}
\rho(A(t)) = \sigma(A(t)) \hspace{5mm} \textrm{for all} \hspace{5mm} A(t)\in\mathcal{R}_{\mathcal{O}_2}.\nonumber
\end{eqnarray}

\noindent In contrast, from the perspective of an observer who has access to the observable (projection) $P = \mathbb{I}_1\otimes \vert +\rangle_2\langle+\vert \otimes\mathbb{I}_3\in\mathcal{R}_2$, the states $\rho$ and $\sigma$ \textit{can} be distinguished at any time $t>0$. In other words, this observer interprets the state $\sigma$ as a non-zero probability measure for the excitation of system $2$.  But the projection $P$ is a nonlocal observable (an element of type I factor), to which the observer localised in $\mathcal{O}_2$ has no access, and thus in their eyes the probability for the excitation of system $2$ remains zero during $t<d/c$.\\

\noindent Type III factor, as a mathematical framework for describing local quantum phenomena, has two important implications: $(i)$ the unavoidable presence of entanglement in the global state across separated regions of space time, and $(ii)$ the existence of an intrinsic notion of dynamics - the modular structure - encoded in the algebra of local observables.\\

\noindent A remarkable consequence of locality in quantum systems, as represented in the language of type III von Neumann algebra, is that any global state with bounded energy is entangled across regions of space time. More precisely, this is an implication of the Reeh-Schlieder theorem \citep{RS}, which states that any global state with bounded energy is \textit{cyclic}  for \textit{every} local algebra. The cyclic property means that the action of the local algebra $\mathcal{R}_{\mathcal{O}}$ on any global state $\Psi$ with bounded energy, can generate any other state on the entire Hilbert space to an arbitrary accuracy:
\begin{eqnarray}
\overline{\mathcal{R}_{\mathcal{O}}\vert \Psi\rangle} = \mathcal{H}. \nonumber
\end{eqnarray} 
\noindent Reeh-Schlieder theorem implies that by acting on a state such as vacuum $\Omega$ by an operator located in a bounded region $\mathcal{O}$, one can generate an arbitrary influence outside $\mathcal{O}$ in the causally disconnected region $\mathcal{O}^{\prime}$, which is described by some excited state $\Psi$, $\mathcal{R}_{\mathcal{O}}\vert \Omega\rangle = \vert \Psi\rangle$. To illustrate the operational implication of this, let us imagine the observable $A^{\prime}\in\mathcal{R}_{\mathcal{O}^{\prime}}$ that has zero expectation for the vacuum state $\Omega$, and unit expectation for the excited state $\Psi$ \textit{e.g.} $A^{\prime}$ can encode the probability for the existence of some particle or some complex system in the region $\mathcal{O}^{\prime}$ :  $\omega(A^{\prime}) \equiv \langle\Omega\vert A^{\prime}\vert \Omega\rangle = 0$ and $\psi(A^{\prime}) \equiv \langle\Psi\vert A^{\prime}\vert \Psi\rangle = 1$. But according to the Reeh-Schlieder theorem, $A\vert \Omega\rangle = \vert \Psi\rangle$ for some local operator $A\in \mathcal{R}_{\mathcal{O}}$, which leads to the equation:
\begin{eqnarray}
\omega(A^{\ast}AA^{\prime}) = \psi(A^{\prime}) = 1
\label{commutative}
\end{eqnarray}	
\noindent where we have used the commutativity between the local algebras of space-like separated regions $\mathcal{R}_{\mathcal{O}}$ and $\mathcal{R}_{\mathcal{O}^{\prime}}$ : $[A,A^{\prime}] = 0$. One can clearly see that the global state $\omega(A^{\ast}AA^{\prime})$ can not be factored in terms of the product of the local states - the restriction of the global state to the regions $\mathcal{O}$ and $\mathcal{O}^{\prime}$: $\omega(A^{\ast}AA^{\prime})\neq \omega(A^{\ast}A)\omega(A^{\prime})$,
for otherwise, this would violate the Eq.(\ref{commutative}) since $\omega(A^{\prime}) = 0$.  This means that the pair of commuting observables $\{A^{\ast}A$ , $A^{\prime}\}$, are \textit{correlated} in the vacuum state. 
It turns out that these observable pair correlations are symptomatic of a much profound property linked with the vacuum state, namely, the entanglement \citep{clifton, RevModPhys.90.045003} across the entire region  $\mathcal{O}$ and its causal complement $\mathcal{O}^{\prime}$: 
\begin{equation}
\omega(AA^{\prime}) \neq \omega_1(A)\omega_2(A^{\prime}) \hspace{5mm}\textrm{for all}\hspace{5mm}A\in\mathcal{R}_{\mathcal{O}}, A^{\prime}\in\mathcal{R}_{\mathcal{O}^{\prime}}. \nonumber
\end{equation}
To see whether the entanglement is unique to the vacuum state or whether it is a generic property of type III structure, let us imagine that there exists a global state that is not entangled across the local algebras. Due to the Reeh-Schlieder theorem, one could then generate any other state (including the vacuum state) in the Hilbert space by acting on this global state with some local operator. However, since entanglement, as a global structure, can not be induced locally \citep{clifton}, all the generated states are likewise not entangled, which contradicts the fact that the vacuum  is an entangled state. An important observation here is that this entanglement is a direct consequence of imposing a local structure upon the algebra of observable. Moreover, the degree of entanglement across adjacent regions such as $\mathcal{O}$ and $\mathcal{O}^{\prime}$ is universally divergent \citep{RevModPhys.90.035007}, and does not depend on any particular state, and as stated by E. Witten \cite{RevModPhys.90.045003}: `This ultraviolet divergence means that the entanglement is not just a
property of the states but of the algebras of observables.'\\

\noindent In this work, we are interested in a certain class of local algebras called \textit{diamonds} that encode the observable world of an observer with a finite lifespan. A diamond is an algebra of observables that are localised within a diamond-shaped region of space time bounded by the intersection of the future light-cone of when the observer is created, and the past light-cone of when the observer is annihilated. Therefore, a diamond represents the world that is accessible to a finite observer for measurements. An important property of the diamond $\mathcal{R}_{\diamond}$ is that the entanglement across $\mathcal{R}_{\diamond}$ and $\mathcal{R}_{\diamond^{\prime}}$ can not be broken by local operations such measurements \citep{clifton}. The operational meaning of this is that the local observer can never disentangle their observable world from the rest of the universe by measurement, and therefore, form the perspective of the observer, the world always appears in a mixed state with intrinsic uncertainty.\\

\noindent A mathematically rigorous machinery in which one can analyse the structure of the  algebras that are intrinsically entangled is called the \textit{modular theory} or the \textit{Tomita-Takesaki} theory \citep{TT}. The modular theory entails the existence of a notion of dynamical symmetry that is inherited in the structure of the local algebra.	This intrinsic symmetry was discovered through the investigation of a curious mismatch between the algebraic structure of the observables on one hand, and the induced geometric structure of the Hilbert space on the other. More concretely, the isometry induced by the adjoint operation of the local algebra: $\mathcal{S}_{\mathcal{O}}: A\rightarrow A^{\ast}$ is broken  by the geometric structure of the generated Hilbert space:  $\mathcal{S}_{\Psi}: A\vert \Psi\rangle\rightarrow A^{\ast}\vert \Psi\rangle$, where $\Psi$ is a global state with bounded energy \textit{i.e.}, $\Psi$ is a cyclic and separating\footnote{A state $\Psi$ is called separating for $\mathcal{R}_{\mathcal{O}}$ if $A\vert \Psi\rangle\neq 0$ unless $A$=0 for all $A\in\mathcal{R}_{\mathcal{O}}$.} state of the local algebra $\mathcal{R}_{\mathcal{O}}$. The operation $\mathcal{S}_{\Psi}$ can be broken down in terms of a pair of operators:
\begin{eqnarray}
\mathcal{S}_{\Psi} = J_{\Psi}\Delta_{\Psi}^{1/2}
\label{modularoperator}\nonumber
\end{eqnarray}
where $J_{\Psi}$ is an anti-unitary operator, and $\Delta_{\Psi}$ is a self-adjoint positive operator. It is the existence of the nontrivial operator $\Delta_{\Psi}$, the \textit{modular operator}, which breaks the isometry in the structure of the Hilbert space, for $J_{\Psi}$ can simply be viewed as a reflection around a certain point, and thus it does not induce any distortion in the metric structure of the Hilbert space. The pair $\{J_{\Psi}, \Delta_{\Psi}\}$ encode invaluable information about the structure of the algebra, on which the modular theory is based. The essence of the modular theory is that the modular operator $\Delta_{\Psi}$ gives rise to a symmetry, \textit{the unitary group of automorphisms $\alpha_t^{\Psi}$} under which the local algebra $\mathcal{R}_{\mathcal{O}}$ remains invariant:
\begin{equation}
\alpha_t^{\Psi} [\mathcal{R}_{\mathcal{O}}] \equiv \Delta_{\Psi}^{it}\mathcal{R}_{\mathcal{O}}\Delta_{\Psi}^{-it} = \mathcal{R}_{\mathcal{O}} \hspace{3mm},\hspace{3mm} t\in \mathbb{R} .\nonumber
\end{equation}
In other words, $\alpha_t^{\Psi}$ is a global unitary transform, the restriction of which to the space time region $\mathcal{O}$ preserves the associated local algebra $\mathcal{R}_{\mathcal{O}}$. It is important to note that the modular operator $\Delta_{\Psi}$ contains information of both the local algebra and the global state $\Psi$, and thus as a global operator, it is external to any region of space time. The operational implication of this symmetry is that the state of the world from the perspective of a local observer who senses the world through $\mathcal{R}_{\mathcal{O}}$ remains unchanged under the global unitary transform $\alpha_t^{\Psi}$. \\

\noindent As we will see, the modular automorphism gives rise to the ``time flow" of the observable algebra, which is generated from within the algebraic structure - the modular structure. Put differently, the observable algebra, because of its modular structure, is exposed to this dynamical symmetry that is generated intrinsically by $\log \Delta_{\Psi}$\footnote{$\log \Delta$ is called the \textit{modular Hamiltonian}.}.  Finally, we would like to refer to an important result in the modular theory of type III structures due to Connes \cite{connesNG}: a set of two modular automorphism $\{\alpha_t^{\Psi}, \alpha_t^{\Phi}\}$ induced by the same local algebra $\mathcal{R}_{\mathcal{O}}$, but by two distinct states $\{\Psi,\Phi\}$ are \textit{inner equivalent}. Inner equivalence simply means that the two modular automorphisms are equivalent up to a unitary transform $u \in \mathcal{R}_{\mathcal{O}}$. To understand the implication of this equivalence one should note that the modular automorphisms are global operators, and can not be solely generated from the local algebra as $\Delta \not\in\mathcal{R}_{\mathcal{O}}$, but instead are induced by both the local algebra and some global state. The inner equivalence however, implies that this global state is arbitrary, and that modular flow is state-independent up to a unitary equivalence. We see once again that imposing a local structure on the quantum mechanical description of the world leads to the formation of an exotic class of algebras - Type III factors - that, independently of any particular state,  are intrinsically dynamical. As stated by Connes \cite{connes}: `This flow
\textit{is canonical: it depends only on the algebra itself. Von Neumann algebras, indeed, are}
classified by studying this canonical flow.'

\section{Time Phenomenon: An Epistemic Analysis}

In this section we provide an epistemic analysis of time in terms of the observer's capacity in distinguishing the states corresponding to distinct moments in a world that is unitary and local.\\

\noindent As we saw in the section 2, relative entropy encodes the distinguishability of two distinct states from the perspective of an observer. Eq.(\ref{RE}) however, is valid only in a non-local quantum mechanical world, and thus it is a special case that applies to type I factors only. A more suitable expression for relative entropy that applies to any class of von Neumann algebra was pioneered by Araki \cite{araki} :
\begin{eqnarray}
H_{\Psi,\Phi} = -\langle\Psi\vert \log\Delta_{\Psi,\Phi}\vert \Psi\rangle
\label{RE2}.\nonumber
\end{eqnarray}
Here $\Delta_{\Psi,\Phi}$ is called the \textit{relative} modular operator induced by the adjoint operation:
\begin{eqnarray}
\mathcal{S}: A\vert \Psi\rangle \rightarrow A^{\ast}\vert \Phi\rangle\hspace{5mm}, \hspace{5mm}S_{\Psi,\Phi} = J_{\Psi,\Phi}\Delta_{\Psi,\Phi}^{1/2} \nonumber
\end{eqnarray}
with $A\in\mathcal{R}_{\mathcal{O}}$. $H_{\Psi,\Phi}$ can be construed as a measure of the confidence with which the observer who is located in the region $\mathcal{O}$ - sensing the world through measuring $\mathcal{R}_{\mathcal{O}}$ - can distinguish the true global state $\Psi$ from the expected global state $\Phi$. An important observation here is that the relative entropy involves the logarithm of the  modular operator $\Delta_{\Psi,\Phi}$. It is the existence of this modular operator as a nontrivial object that makes it possible for an observer to perceive distinctions across states. In other words, if the von Neumann algebra did not admit a modular structure ($\Delta = \mathbb{I}$), then $\log \Delta  = 0$, and thus all the possible distinct states of the world would have appeared indistinguishable in the eyes of the observer, rendering the conception of dynamics impossible. This is a manifestation of the inextricable link between the modular structure and the intrinsic dynamism of the type III von Neumann algebra.\\

\noindent Furthermore, The monotonicity of relative entropy states \citep{araki} that the distinguishability never increases as the region of access $\mathcal{O}$ decreases in size\footnote{Note that the relative entropy $H_{\Psi;\Phi}$ is a function of the local algebra $\mathcal{R}_{\mathcal{O}}$, as the modular operator is generated by both the global state and the local algebra.}:
\begin{eqnarray}
H_{\Psi,\Phi}^{\mathcal{O}} \leq H_{\Psi,\Phi}^{\tilde{\mathcal{O}}} \hspace{5mm},\hspace{5mm} \mathcal{O}\subseteq\tilde{\mathcal{O}} \nonumber
\end{eqnarray}
That is, the larger the set of observables with witch the observer senses the world, the better they can distinguish the global states. To obtain an intuitive understanding of the monotonicity relation one should note that in the type III structures, in particular the diamond algebras, due to the unavoidable presence of entanglement in the global state, the local observer will always see a highly mixed state with high degree of uncertainty. As the observer's region of access for measurements decreases, the uncertainty that is inherent in the observed state increases, and obscures the distinctiveness of the global states even further.\\

\noindent The most primary aspect of temporality is the perceived distinctiveness of the world-states corresponding to separate moments in the eyes of the observer. To see how this phenomenon can be mathematically represented in terms of the relative entropy, we note that $H_{\psi,\phi}$ is a temporal object in that it captures (at the moment of observation and from the perspective of the observer) the distinction between the  state $\psi$ that is \textit{present} and yet to be measured, and the expected state $\phi$ that simply exists as attained information or as a belief that must have occurred to the observer through a \textit{prior} measurement of some other state $\tilde{\psi}$ that is \textit{absent}. Therefore, the distinction between the states of the world $\{\psi , \tilde{\psi}\}$ corresponding to separate moments can be viewed as the variation in the amount of attainable information through measuring the states $\{\psi , \tilde{\psi}\}$. That is, the variation in the amount of attainable information quantifies the degree of relative surprise upon measuring the state of the world at one moment with respect to that of the previous moment. Therefore, the distinction between the states $\{\psi , \tilde{\psi}\}$ as expressed in terms of the information gain reads:
\begin{equation}
\mathfrak{S}_{\psi, \tilde{\psi}} \equiv \sup \chi(\{\psi_x; p_x\}) - \chi(\{\tilde{\psi}_x; p_x\}).\hspace{5mm} \nonumber
\label{QP}
\end{equation}
Here $\chi(\{\psi_x; p_x\})$ is called the Holevo bound \citep{qit}, and it captures the upper bound on the amount of attainable information upon measuring the state $\psi = \sum_xp_x\psi_x$, which can be expressed \citep{RevModPhys.74.197} in terms of the relative entropy between $\psi$ and the measuring basis $\{\psi_x\}$: 
\begin{equation}
\chi(\{\psi_x; p_x\}) = \sum_x p_xH_{\psi_x, \psi}. \nonumber
\end{equation}
The supremum is over all possible set of basis $\{\psi_x\}$. One can see that if the states $\{\psi , \tilde{\psi}\}$ are indistinguishable, then there will be no variation in the information gained upon measuring the states. Alternatively, for an observer who can not distinguish $\psi$ from $\tilde{\psi}$, there will be no surprise upon measuring $\psi$ having already measured $\tilde{\psi}$, and thus the observer will not perceive any change.\\

\noindent Let us highlight two important aspects in regards to the interpretation of the Holevo bound: first, $\chi(\{\psi_x; p_x\})$ encodes how well the observer can recover $\{p_x\}$ by performing measurement on the state $\psi$ using the basis $\{\psi_x\}$. More concretely, the information contained in $\psi$ is indeed encoded in the probability distribution $\{p_x\}$, and thus $	\mathfrak{S}_{\psi, \tilde{\psi}}$ quantifies how distinct the \textit{retrieved} probability distributions appear in the course of a sequence of measurements. Second, if the basis $\{\psi_x\}$ are perfectly distinguishable from each other -  $\{\psi_x\}$ are all pure and orthogonal - then the Holevo bound simply reduces to the entropy of the state $\psi$. That is, the observer can know \textit{all} there is to know that is encoded in the state $\psi$. This is always the case in classical systems as the measurement basis \textit{e.g.}, two sides of a coin, are perfectly distinguishable. In quantum systems however, the measuring basis can overlap, $\langle \psi_x, \psi_ y\rangle \neq 0$, which results in the decay of the observer's accuracy in retrieving $\{p_x\}$, and that consequently, leads to a partial knowledge  of the state $\psi$. Also, as one would expect, for a pure state the Holevo bound is simply zero as the entropy is zero, and thus there is no information to be attained.\\

\begin{figure}[!htb]
	\hspace{10mm}\includegraphics[scale=0.35]{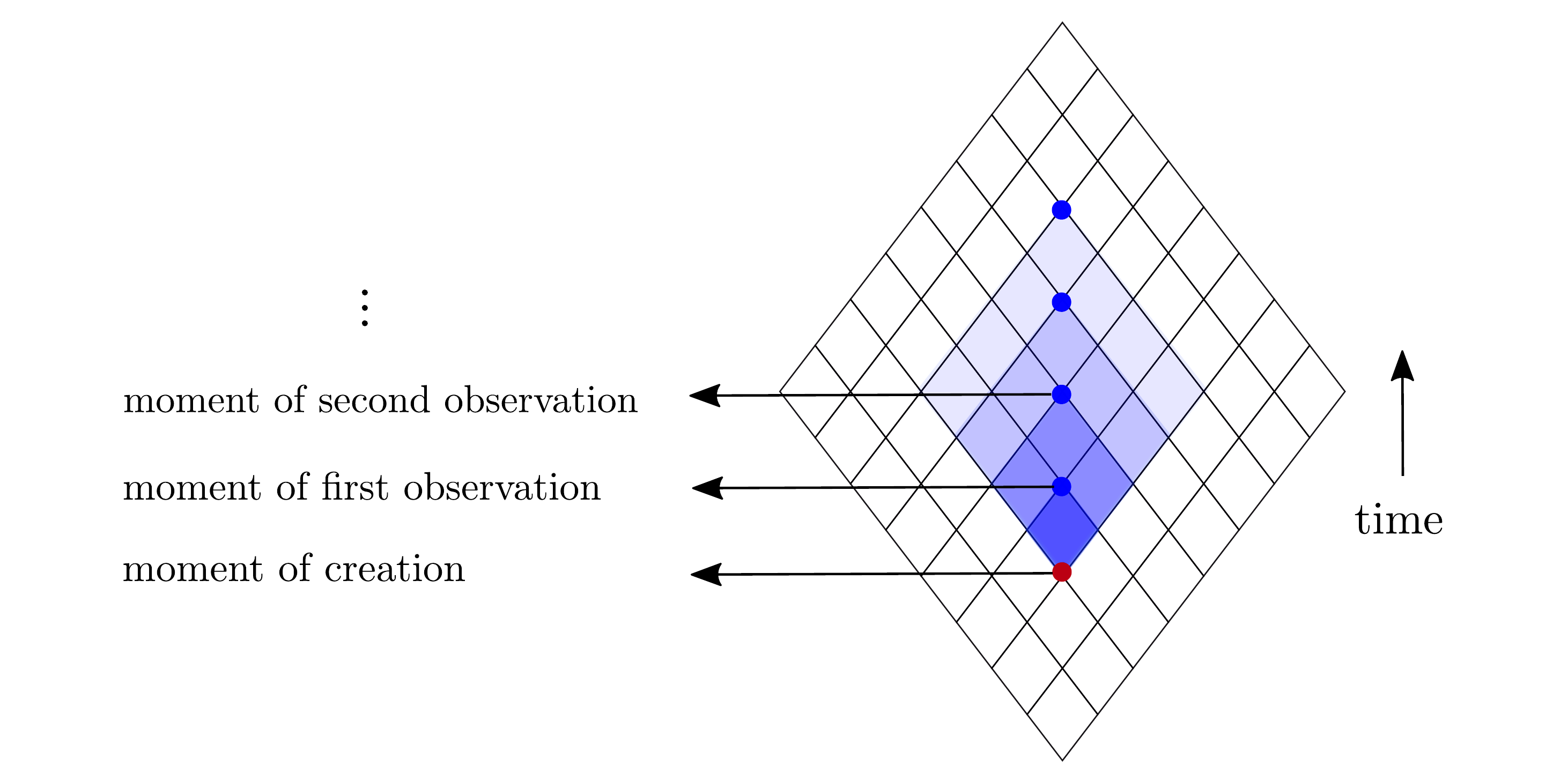}
	\caption{The depiction of a sequence of nested diamond algebras each representing the observable world of a finite observer at each moment of their life.  }
	\label{diamond}
\end{figure}

\noindent Now let us suppose that the observer is thrown into the world that is described in terms of the type III algebra. At each moment of observation, the observer senses the world from the perspective of a diamond algebras $\mathcal{R}_{\diamond}$ that is nested inside a larger algebra, through which the next observation is made (see figure \ref{diamond}.)
Due to the unavoidable presence of entanglement in the global state $\Psi$ across the diamond algebras, at each moment of observation, the observer, who is confined to a diamond, is measuring a highly mixed state $\psi^{\diamond} = \langle\Psi\vert \mathcal{R}_{\diamond}\vert \Psi\rangle$. Thus, from the perspective of the observer, the distinction between the mixed states $\psi^{\diamond}$ and $\psi^{\tilde{\diamond}}$ with $\tilde{\diamond} \subset \diamond$ can be written as the variation in the information that can be attained from measuring the mixed states: 
\begin{equation}
\mathfrak{S}_{\diamond, \tilde{\diamond}} = \sup \chi(\{\psi^{\diamond}_x; p_x\}) - \chi(\{\psi^{\tilde{\diamond}}_x; p_x\})
\label{CRE}
\end{equation} 
where the state $\psi^{\diamond} = \sum_xp_x\psi^{\diamond}_x$, is measured in the basis states $\{\psi^{\diamond}_x\}$. The definition (\ref{CRE}) is called the Connes-St\o rmer relative entropy \citep{connesEntropy}, and it can be construed as a measure of how distinct a global state $\Psi$ can appear when it is measured from the perspective of a pair of nested local algebras  $\mathcal{R}_{\mathcal{O}}$ and $\mathcal{R}_{\tilde{\mathcal{O}}}$ with $\tilde{\mathcal{O}} \subset \mathcal{O}$. In the field of quantum information theory, $\mathfrak{S}_{\mathcal{O}, \tilde{\mathcal{O}}}$ is sometimes referred to as the 
\textit{quantum privacy} \citep{quantumprivacy} as it quantifies the amount of information that is in principle hidden or invisible to the observer who is confined to the sub-algebra $\mathcal{R}_{\tilde{\mathcal{O}}}$.\\

\noindent Due to the monotonicity of relative entropy, $H_{\psi^{\tilde{\diamond}}_x, \psi^{\tilde{\diamond}}}< H_{\psi^{\diamond}_x, \psi^{\diamond}}$,  $\mathfrak{S}_{\diamond, \tilde{\diamond}}>0$. That is, the observer's partial knowledge increases each time that they perform a measurement. For the observer to perceive perfect distinctiveness however, the variation in the attained information should diverge. That is for the two states $\{\psi^{\diamond}, \psi^{\tilde{\diamond}}\}$ to appear perfectly distinguishable, the corresponding  Connes-St\o rmer relative entropy should be infinite. Intuitively, the variation in the attainable information as the observer's local algebra grows is proportional to the relative `size' of the algebras. That is, the larger the ratio of the local algebras $\{\mathcal{R}_{\mathcal{O}}, \mathcal{R}_{\tilde{\mathcal{O}}}\}$, the more information to be gained through channeling from one to the other, $\tilde{\mathcal{O}} \rightarrow \mathcal{O}$, and subsequently, the larger  $\mathfrak{S}_{\mathcal{O}, \tilde{\mathcal{O}}}$. A rigorous algebraic measure that quantifies the relative sizes of two sub-factors $\mathcal{R}_{\tilde{\mathcal{O}}} \subset\mathcal{R}_{\mathcal{O}}$ is called the \textit{Jones index} \citep{jones}, $[\mathcal{R}_{\mathcal{O}},\mathcal{R}_{\tilde{\mathcal{O}}}]$. The Jones index is always positive and equals to one if and only if $\mathcal{R}_{\mathcal{O}}=\mathcal{R}_{\tilde{\mathcal{O}}}$. It turns out that for the diamond sub-factors - the sequence of nested algebras representing the observable world of a finite observer - the Jones index is infinite:
\begin{equation}
[\mathcal{R}_{\diamond}, \mathcal{R}_{\tilde{\diamond}}] = \infty \hspace{5mm},\hspace{5mm}\mathcal{R}_{\tilde{\diamond}} \subset \mathcal{R}_{\diamond}.\nonumber
\end{equation} 
Therefore, the global state $\Psi$ always appears distinct at each moment that the observer performs a measurement. \\

\noindent It is of central importance to note that in our analysis, the phenomenon of time - the epistemic distinctiveness - which emerges as a result of observing the world from the perspective of various local algebras, appears to be  insensitive to the ontic state of the world $\Psi$. In other words, any global state, in a world that is unitary and local, appears distinct each time that it is being observed by a local observer. Furthermore, it follows that if this epistemic distinction is independent from the ontic global state, it must as well be independent of any ontic evolution of the global state, since any other state induces the same phenomenon. Put differently, in a unitary and local world, the observer experiences time whether the ontic state of the world is evolving or whether it is invariant. Indeed, by committing to a type III structure as the ontology of our analysis, the global state of the world remains invariant according to the modular structure, $\Delta_{\Psi}^{it}\vert \Psi\rangle =\vert\Psi\rangle$.

\section{Epistemic Analysis of Time and Thermal Time Hypothesis}

\begin{figure}[!htb]
	\hspace{25mm}\includegraphics[scale=0.25]{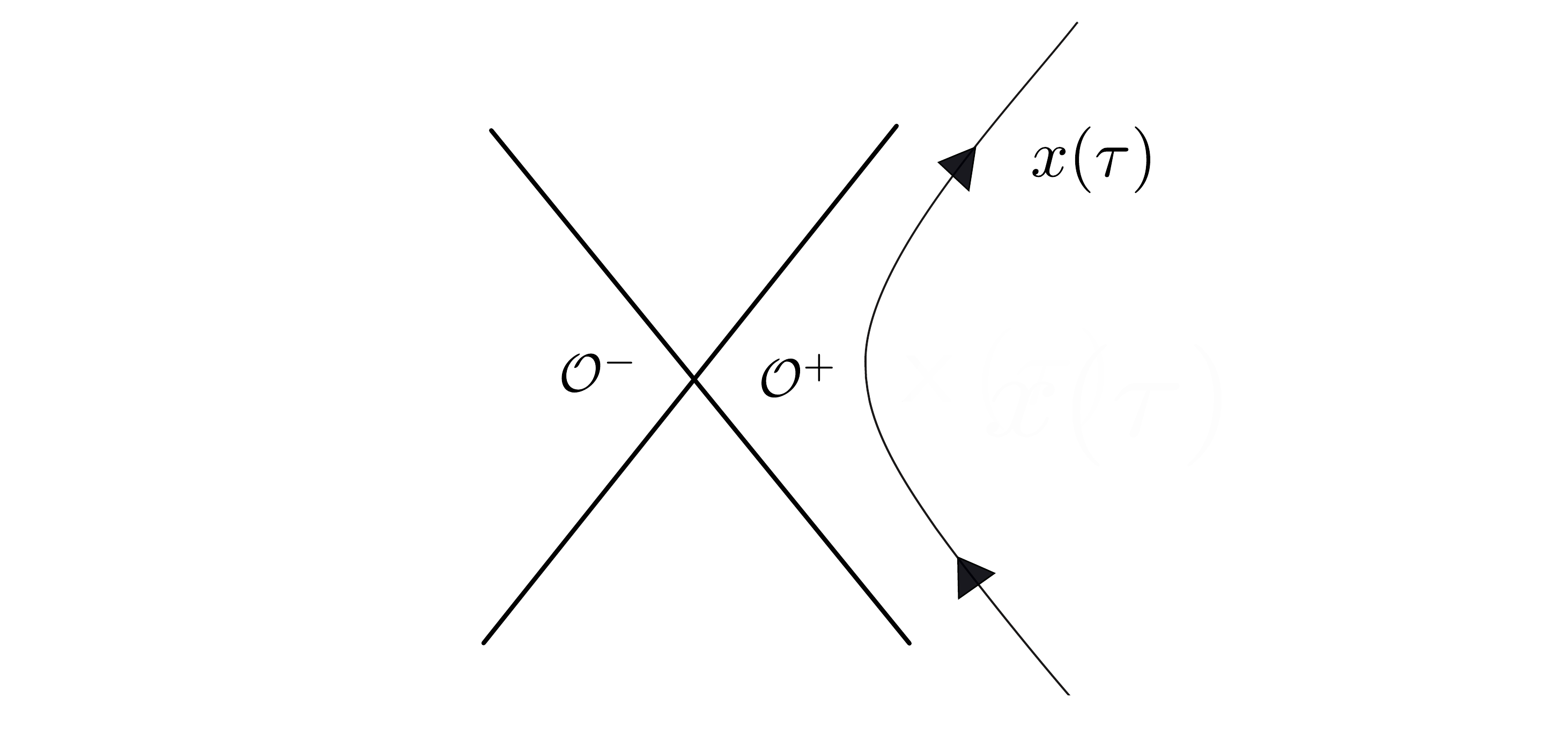}
	\caption{The graph of a uniformly accelerating observer who is restricted, as a consequence of their uniform acceleration, to the right-side region - the Rindler wedge - of the Minkowski space time.}
	\label{TTH}
\end{figure}

Thermal Time Hypothesis \citep{connes}, as a solution to the problem of time in generally covariant quantum theories, is the proposition that time has a thermodynamical origin. The hypothesis, based on the modular structure of the von Neumann algebra, suggests that the unitary group of modular automorphism $\alpha^{\Psi}_t$, that is generated by both the global state $\Psi$ and the local algebra $\mathcal{R}_{\mathcal{O}}$, determines the time flow experienced by the observer localised in the region $\mathcal{O}$. To illustrate this, and more importantly, to understand from where does the term `thermal' originate let us consider the world to be in the vacuum state $\Omega$, in which an immortal observer is moving with the uniform acceleration $a$ (see figure \ref{TTH}.)
The time flow experienced by the this observer can be computed in terms of the unitary transform $u_{\tau} = e^{i a \tau K}$ generated by the boost operator $K$. Due to their uniform acceleration, the observer is restricted to the right-side region - the Rindler wedge - of the space time, and thus they sense the world through the corresponding local algebra $\mathcal{R}_{\mathcal{O}^{+}}$. It turns out that the modular flow induced by the local algebra $\mathcal{R}_{\mathcal{O}^{+}}$ is closely related to the proper time flow of the observer \citep{BW}:
\begin{equation}
\alpha_t [\mathcal{R}_{\mathcal{O}^{+}}] = u_{-\beta t}\mathcal{R}_{\mathcal{O}^{+}}u_{\beta t} .
\label{BW}
\end{equation}

\noindent The significance of the Eq.(\ref{BW}) is that by solely analysing the modular structure of the local algebra $\mathcal{R}_{\mathcal{O}^{+}}$, one is able to deduce the proper time experienced by an observer who is moving with a uniform acceleration. As one can see the proper time and the modular time are related via a proportionality constant , $t = -\beta \tau$, where $\beta = 2\pi/a$.\\

\noindent In order to understand what does the experience of the uniformly accelerating observer consist of, one should note that due to the inherent entanglement of the type III algebra, the vacuum state appears highly mixed from the perspective of the observer localised in the region $\mathcal{O}^{+}$. In fact it can be shown \citep{RevModPhys.90.045003} that the restriction of the vacuum state to the local algebra, $\omega(\mathcal{R}_{\mathcal{O}^{+}}) = \langle\Omega\vert \mathcal{R}_{\mathcal{O}^{+}}\vert \Omega\rangle$, gives rise to a maximally mixed state that is in \textit{thermal} equilibrium with respect to the proper time flow. Indeed, this is a consequence of the modular structure of the local algebra: every local algebra induces a modular flow with respect to which the local state - the restriction of the global state to that local algebra - is in thermal equilibrium. Therefore, the modular flow induced by $\mathcal{R}_{\mathcal{O}^{+}}$ determines the time experience of the observer localised in $\mathcal{O}^{+}$, which consists of a perpetual exposure to a thermal state at the inverse temperature $\beta = 1/T$. The thermal time hypothesis is a general proposition that the proper time of the observer localised in \textit{any} region of space time $\mathcal{O}$, is determined by the local thermal state, as its modular flow.\\

\noindent The thermal picture of time appears to face a conceptual issue concerning the epistemology of the observer. The perpetual perception of a thermal state  renders any epistemic access to time impossible as the situated observer simply does not realise any distinction. The non-equilibrium phenomena that appear to underly the conception ot time itself, do not seem to be addressed in a picture where the primary objects are the equilibrium states. Connes and Rovelli \cite{connes} do not see this as a conceptual issue by observing that in the infinite dimensional quantum systems one typically measures small perturbations around a background thermal state. Therefore, the local observer is not frozen in time, and instead perceives change in the state of affairs due to small fluctuations around an equilibrium state that determines the local proper time as its modular flow. In this view, the background thermal state exists as a limit state that is reached asymptotically, and thus the associated modular flow simply describes the \textit{near equilibrium} behaviour \citep{longoNE}.\\

\noindent An instance where the proper time for a near equilibrium phenomenon is determined by the background thermal state, is presented in the context of Friedman-Robertson-Walker (FRW) cosmology. It has been shown \citep{rovelli} that in a FRW universe that is filled with radiation, the cosmological time - the proper time experienced by co-moving observers - coincides with the flow determined by the Cosmic Microwave Background (CMB) as the background thermal state of the universe. In this analysis, the thermal time can explain the emergence of the cosmological time experienced by the co-moving observers to whom the universe appears isotropic. Swanson \cite{swanson} notes this problem by pointing out the that in our human scale however, the universe appears highly non-isotropic, and thus our experienced time flow, which occurs at much smaller scales, does not seem to be reducible to the thermal time induced by the CMB. Therefore the conceptual challenge facing the thermal time hypothesis, is linked with the time flow in strictly non-equilibrium phenomena that we perceive as local observers.

\begin{figure}[!t]
	\hspace{5mm}\includegraphics[scale=0.35]{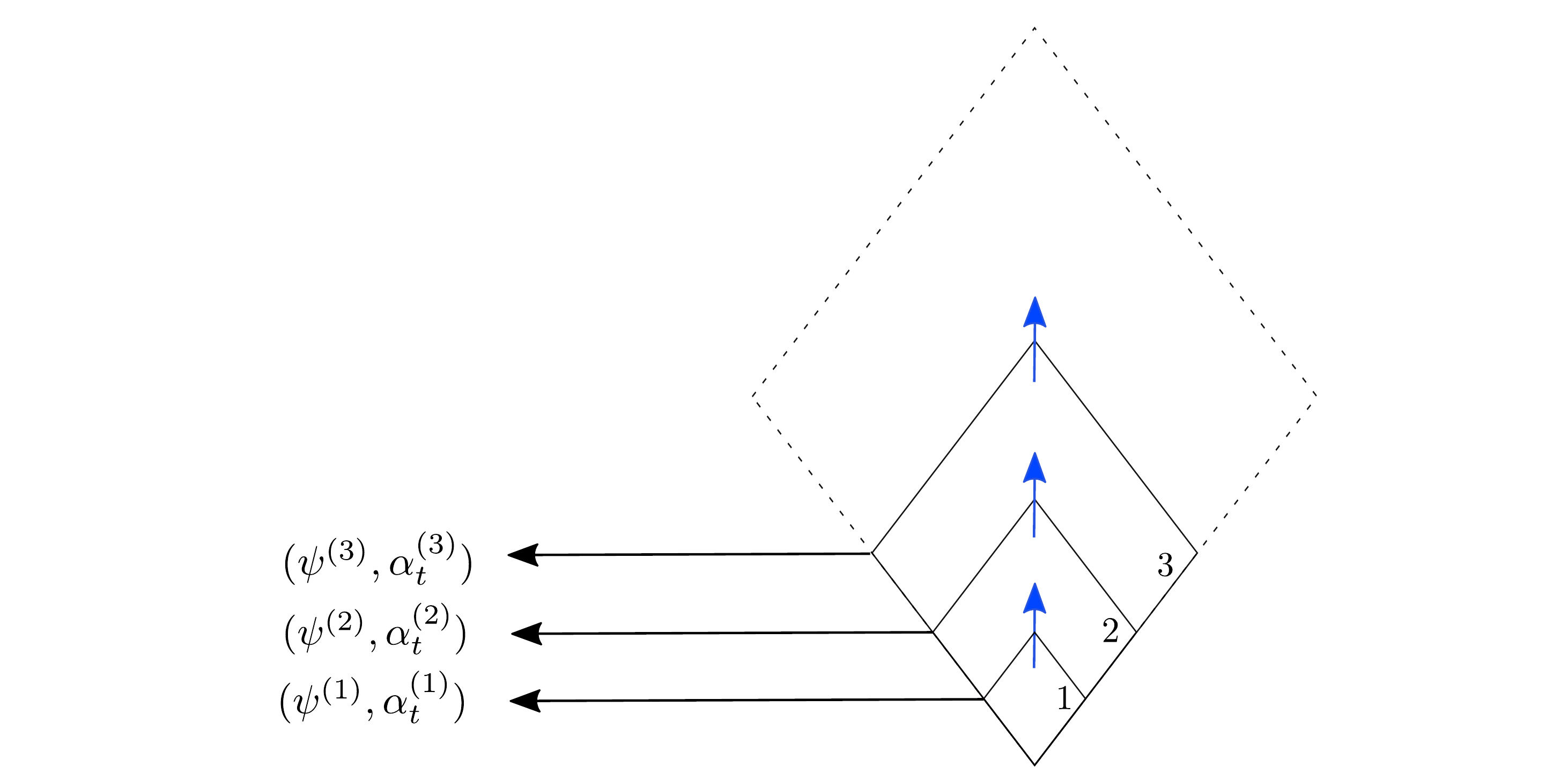}
	\caption{The depiction of the observer's sequential exposure to a series of distinct states that are each at thermal equilibrium with respect to their corresponding modular flows. }
	\label{diamond2}
\end{figure}

\noindent The problem of time in non-equilibrium phenomena seems to be dissolved within the epistemic analysis that we have presented in this work. To see this, we note that at each moment of observation, the finite observer senses the world from the perspective of a local algebra that is slightly different from the previous moment (see figure \ref{diamond}.) Therefore, the observer is never restricted to a single diamond algebra, and instead is \textit{channeling} from one region $\tilde{\diamond}$ to a slightly larger region $\diamond$. In this way, the experienced time of the observer is not determined by the modular flow of some local algebra $\mathcal{R}_{\diamond} $, through which the observer senses the world only once in their lifetime. But instead, the time flow of a finite observer consists of a sequential exposure to a series of distinct local states $\psi^{(i)} = \langle\Psi\vert\mathcal{R}_{\diamond^{(i)}}\vert\Psi\rangle$, that are each in a thermal equilibrium with respect to their corresponding (distinct) modular flows $\alpha^{(i)}_t$ induced by the global state $\Psi$ and the local algebra $\mathcal{R}_{\diamond^{(i)}}$ (see figure \ref{diamond2}.) In other words, the relative entropy $\mathfrak{S}_{\tilde{\diamond},\diamond}$ , captures in the eyes of the observer, the distinction  between a sequence of thermal states $\psi^{(i)}$.  Therefore, what appears to encode the flow of time for an observer, must be a mathematical map that functions as a channel across the sequence of nested local algebras.\\

\noindent We note that the scenario of the immortal observer who is moving with a uniform acceleration is simply a special case of the framework presented here. For an immortal observer, the accessible algebra, $\mathcal{R}_{O^{+}}$, is an infinite diamond that can not be nested within a larger enclosing one. Alternatively, for an immortal observer, the sub-algebra and the algebra coincide, which results in the vanishing of the corresponding Connes-St\o rmer relative entropy $\mathfrak{S}_{\diamond^{\infty}, \diamond^{\infty}} = 0$. Therefore, the observer can not distinguish the thermal states corresponding to separate moments, and thus they perceive no change.

\section{Conclusion}

In this work we have presented an epistemic analysis of time where the distinctiveness perceived by the observer is the primary function that is rigorously represented by the information-theoretic concept of relative entropy. The ontic commitment to a unitary and local world in which the observer is situated, leads to the formation of the algebraic language of type III within which the phenomenon of time emerges as an epistemic function that is in principle insensitive to any particular ontic state of the world. \\

\noindent The analysis presented here can be further explored from a variety of perspectives. For instance, the properties of the algebraic map that functions as a channel across local algebras can be further analysed from the perspective of the modular theory. There exists a class of maps across diamond algebras called the \textit{canonical endomorphisms} \citep{longoCE} that are unitary maps determined by the modular structure. Due their unitary property however, they can not account for the monotonicity of the relative entropy, as $\mathcal{H}_{\Psi, \Phi}$ is invariant under unitary transforms.\\

\noindent A second direction of research consists in addressing the question of how the arrow of time emerges from our analysis. It is not clear how the epistemic arrow of time emerges within the presented framework. That is, the arrow of time, which can be be thought of as the tendency towards equilibrium, entails a reduction in the observer's capacity to distinguish the state of affairs. This reduction however is not visible in our analysis as the relative entropy between states at consecutive moments is infinite due to the divergence of the Jones index.\\

\noindent Finally, a challenging yet intriguing attempt is to explore the possibility of the emergence of the locality as an epistemic function. In this line of research, the goal would be to minimise our ontic commitments by bracketing the local structure of the world from our analysis, and examine  whether spacetime itself, as a local background, can be grounded in the perception of the observer who is situated in a world described by a simpler ontology of type I algebra.

\section*{Declarations}

	 {\bf{Funding}} This research has received funding from the European Union’s Horizon 2020 research and innovation programme under grant agreement n. 758145.\\
	 
	\noindent {\bfseries{Conflict of interest}} The authors do not have any relevant financial or non-financial interests to disclose.\\

 \noindent {\bfseries{Availability of data and materials}} Not applicable.\\
 
	 \noindent {\bfseries{Code availability}} Not Applicable.


\bibliography{ref}



\end{document}